\documentstyle[12pt]{article}

\def\plb#1 #2 {Phys. Lett. {\bf#1B} #2 }
\def\npb#1 #2 {Nucl. Phys. {\bf B#1} #2 }
\def\prl#1 #2 {Phys. Rev. Lett. {\bf #1} #2 }
\def\cmp#1 #2 {Comm. Math. Phys. {\bf #1} #2 }
\def\mpl#1 #2 {Mod. Phys. Lett. {\bf A#1} #2 }

\def\inbar{\vrule height1.5ex width.4pt depth0pt}
\def\IB{\relax{\rm I\kern-.18em B}}
\def\IC{\relax\,\hbox{$\inbar\kern-.3em{\rm C}$}}
\def\ID{\relax{\rm I\kern-.18em D}}
\def\IE{\relax{\rm I\kern-.18em E}}
\def\IF{\relax{\rm I\kern-.18em F}}
\def\IG{\relax\,\hbox{$\inbar\kern-.3em{\rm G}$}}
\def\IH{\relax{\rm I\kern-.18em H}}
\def\II{\relax{\rm I\kern-.18em I}}
\def\IK{\relax{\rm I\kern-.18em K}}
\def\IL{\relax{\rm I\kern-.18em L}}
\def\IM{\relax{\rm I\kern-.18em M}}
\def\IN{\relax{\rm I\kern-.18em N}}
\def\IO{\relax\,\hbox{$\inbar\kern-.3em{\rm O}$}}
\def\IP{\relax{\rm I\kern-.18em P}}
\def\IQ{\relax\,\hbox{$\inbar\kern-.3em{\rm Q}$}}
\def\IR{\relax{\rm I\kern-.18em R}}
\def\ZZ{\relax{\sf Z\kern-.4em Z}}

\def\beq{\begin{equation}}    \def\eeq{\end{equation}}
\let\nn=\nonumber
\def\beann{\begin{eqnarray*}} \def\eeann{\end{eqnarray*}}
\def\bea{\begin{eqnarray}}    \def\eea{\end{eqnarray}}
\def\lleq#1{\label{#1}\eeq}   \def\llea#1{\label{#1}\eea}
\def\nn{\nonumber}
\def\fnote#1#2{\begingroup\def\thefootnote{#1}\footnote{#2}\addtocounter
{footnote}{-1}\endgroup}

\def\notin{\ \hbox{{$\in$}\kern-.51em\hbox{/}}}

  \def\th{\theta}  
\def\bz{\bar z} \def\bth{\bar {\theta}}
   
   \def\cI{{\cal I}}
   
 \def\cO{{\cal O}}

\def\tn{{\tilde n}} \def\tW{\widetilde W} \def\tPhi{\widetilde {\Phi}}
\def\tPsi{\widetilde {\Psi}}
\renewenvironment{thebibliography}[1]
  { \begin{list}{\arabic{enumi}.}
    {\usecounter{enumi} \setlength{\parsep}{0pt}
     \setlength{\itemsep}{3pt} \settowidth{\labelwidth}{#1.}
     \sloppy
    }}{\end{list}}

\begin{document}
\hfill {
{HD--THEP--92--53}}
\vskip .05truein
\hfill {

\vskip .4truein
\parindent=1.5pc

\begin{center}{
   {{\bf Marginal Flows between Mirror Pairs of Landau--Ginzburg
                String Vacua: \\
                Thickening Moduli Space Via $c=0$ Theories}
\fnote{\diamond}{Based in part on a talk presented at the Trieste Summer
                 Workshop on Superstrings, Trieste, Italy, 1992}
}\\
\vglue 2cm
{Rolf Schimmrigk}\\ [2ex]
\baselineskip=16pt
{\it Institut f\"ur Theoretische Physik, Universit\"at Heidelberg}\\
{\it Philosophenweg 16, 6900 Heidelberg, FRG}\\
\vglue 4.5cm
{ABSTRACT}}
\end{center}
\vglue 0.3cm
{\rightskip=1.2pc
 \leftskip=1.2pc
\baselineskip=14pt
\noindent
A recently introduced method for constructing marginal singular flows
between distinct Landau--Ginzburg theories at fixed central charge is
reviewed. The flows are constructed in an enlarged moduli space obtained
by adding theories with zero central charge. This mechanism is used to
construct flows between mirror pairs of string vacua described by N$=$2
superconformal Landau--Ginzburg fixed points. In contrast to previous
methods this new construction of mirror theories does not depend on
particular symmetries of the original theory.
\vglue 0.8cm}
\renewcommand\thepage{}
\vfill \eject
{\bf\noindent 1. Introduction}
\vglue 0.2cm
\baselineskip=17.5pt
\pagenumbering{arabic}

\noindent
Most known constructions of mirror pairs of Landau--Ginzburg string vacua
and  Calabi--Yau manifolds depend, in general, on orbifolding of the
original
theory \cite{cls}\cite{ls4}\cite{gp}\cite{bh}
\fnote{1}{The only exception is Batyrev's construction \cite{vitja} of
mirrors
          in terms of dual polyhedra.}.
This is unsatisfactory because no general selection rule is
known which determines the specific form of the action of the symmetry to be
modded out or even the order of the group.
Even though for Fermat type superpotentials both ingredients,
the order of the group and its specific action, are known \cite{unpub}, for
non--Fermat type superpotentials finding the action is more of an art. Thus it
would be useful to find a construction which does not depend on the modding of
symmetries.

In this review I discuss a new construction introduced recently \cite{mine}
which does not depend on the modding of discrete symmetries. The idea is based
on a new type of flow between distinct Landau--Ginzburg theories at fixed
central charge: Consider a heterotic vacuum described by a Landau--Ginzburg
theory defined by a superpotential $W(\Phi_i)$ that depends on chiral
$N=2$ superfields $\Phi_i(z,\bz,\th^{\pm},\bth^{\pm}),~i=1,\dots,n$.
The question is how to construct the mirror theory, which I assume
to be described by the mirror potential $\tW(\tPhi_i)$ which depends on chiral
$N=2$ superfields $\tPhi_i(z,\bz,\th^{\pm},\bth^{\pm}),~i=1,\dots,\tn$.
Because the two potentials $W(\Phi_i),\tW(\tPhi_i)$ are mirrors of each other
they represent two different formulations of the same underlying theory.
A first step toward an explicit construction of the mirror potential is
to assume a candidate pair as given and to somehow `move' from one
 representation to the other, or establish any kind of map between them.

In order to achieve this one might imagine adding the two potentials
and performing some sort of projection in the path integral which reduces
this `superposition' of theories to the two individual representations.
This does not work however because the central charge is doubled in the
process and therefore the theory $W+\tW$ does not describe a heterotic
vacuum of the type necessary. This problem can be circumvented however.

In Section 2 I will describe the idea in the general framework of $N=2$
supersymmetric Landau--Ginzburg theories. Even though
the application will be mirror symmetry, the basic framework is
quite independent of that particular question and may well be of interest
in other contexts. Concrete applications to various types of potentials
will be presented in Section 3 and in Section 4 I will describe flows between
mirror pairs of
Landau--Ginzburg vacua in this framework. The review ends with some remarks
concerning the explicit construction of mirrors in this new framework
in Section 5 without assuming that the mirror pair $(W,\tW)$ is known
a priori.

\vskip .4truein
{\bf\noindent 2.  Thickening Moduli Space via $c=0$ Theories}
\vglue 0.4cm
\noindent
The basic idea is to compensate the central charge of the new theory by
introducing
additional fields with negative central charge
\fnote{1}{It should be emphasized that this does not mean that
          the {\it dimensions} of these additional fields are negative,
          which would be a disaster.}
such that the added theory has in fact vanishing central charge.
Given an arbitrary $N=2$ supersymmetric Landau--Ginzburg theory with
some central charge it is possible to add some new nontrivial theory
without changing the central charge by doing something that I will call
`trivialization'.
To find such fields and the appropriate action is simple and always
possible: for any superpotential $\tW(\tPhi_i)$ the potential
\beq
\tW^{'}(\tPhi_i,\Psi_i) := \tW(\tPhi_i) + \sum_i \tPhi_i\Psi_i
\eeq
leads to a $c=0$ theory. I will call this theory the `trivialized' theory.
Adding this trivialized theory to the original potential $W(\Phi_i)$
therefore does not change the central charge. Furthermore the
spectrum remains unchanged because the chiral ring of the theory $\tW'$ is
trivial owing to the fact that $q_{\Psi_i} = 1 - q_{\tPhi_i}$,  hence
the dimension of the polynomial ring, described by the Milnor number,
becomes
\beq
\mu(\tW')=\prod_i\left(\frac{1}{q_{\tPhi_i}-1}\right)
        \prod_j\left(\frac{1}{q_{\Psi_j}-1}\right) = 1.
\eeq
More explicitly this can be seen by considering the ideal generated
          by the potential $\tW'$ which shows that the order parameters
are all trivial.

Adding the
potential $\tW^{'}(\tPhi_i,\Psi_i)$ is, however, not merely a trivial
rewriting of the original
theory because the parameter space of the new theory can contain deformations
along `mixed' directions, involving fields of the type
$\cO_{IJ}= \Phi_I \Psi_J$, where $I,J$ are multiindices, i.e. the operators
$\cO_{IJ}$ describe monomials in the
variables $\Phi_i$ and $\Psi_j$. The existence of these mixed directions
is nontrivial. Assuming they exist the question arises what their general
structure is. Since the variables with negative central charge were introduced
via terms of the form $\tPhi_i\Psi_i$ these new operators are not bilinear
but instead will be, in general, of the form $\Phi_i^{a_j} \Psi_j$ with
positive integer exponents $a_j>1$. Denote the
potential containing the couplings between the
 $\Phi_i$'s and the $\Psi_j$'s by $ W(\Phi_i,\Psi_j)$. It is then possible to
move to a point in the enlarged parameter space where the total potential
is of the form
\beq
W(\Phi_i) + W(\Phi_i,\Psi_j) + \tW(\tPhi_i),
\eeq
i.e. the $\Psi_j$ are completely decoupled from the $\tPhi_i$. At this point
the theory then becomes more singular than the original theory: the singular
set is not just an isolated point at the origin but some higher dimensional
submanifold. This is
precisely what is needed because it allows the spectrum of the theory to
change.
If it so happens that the first two potentials define a $c=0$ theory it is
possible to split off this new trivialized theory to be left with a potential
that corresponds to a different consistent Landau--Ginzburg theory
with an isolated singularity, i.e. with a finite chiral ring.

In a nutshell, then, the idea is to add new directions to the moduli space
of Landau--Ginzburg theories by adding $c=0$ theories. Moving in this enlarged
configuration space allows to connect theories with different spectra by
passing through degenerate theories which can be made into regular theories
by splitting off different $c=0$ theories.
\vglue 0.6cm
{\bf\noindent 3. Examples: Coset Landau--Ginzburg Theories and Others}
\vglue 0.4cm
\noindent
To be concrete consider first a pair of potentials which derives
\cite{m}\cite{vw} from a minimal
$N=2$ superconformal theory at level $k=6$ with central charge $c=9/4$; the
first one
\beq
W(\Phi_i) = \Phi_1^8
\lleq{min6a}
describes the mean field theory of the exactly solvable model endowed with the
diagonal affine invariant whereas the second one
\beq
\tW(\tPhi_i) = \tPhi_1^4+\tPhi_1\tPhi_2^2
\lleq{min6d}
describes the theory with the nondiagonal affine D--invariant.

The extended theory
\beq
\tW'(\tPhi_i,\Psi_i) =
\tPhi_1^4+\tPhi_1\tPhi_2^2 +\tPhi_1\Psi_1+\tPhi_2\Psi_2.
\eeq
is indeed trivial as can be seen from the ideal $\cI$ \beq
\cI =\left[4\tPhi_1^3+\tPhi_2^2+\Psi_1,~2\tPhi_1\tPhi_2+\Psi_2,~
           \tPhi_1,~\tPhi_2 \right]
\eeq
which is generated by the partials of the superpotential $W$. Thus the
physical spectrum of this $c=0$ theory indeed just consists of the vacuum.
It should also be noted that because the theory is transverse except
at the origin,
no new singularities are introduced by adding this $c=0$ theory to
the original potential.

The enlarged moduli space of this model does contain mixed monomials of
charge one, e.g.
\beq
\Phi_1^2\Psi_1,~~~~ \Phi_1^3\Psi_2
\eeq
which introduce new marginal directions along which the original theory
can be deformed.  Thus it is possible to move to points in parameter space,
 defined by
\beq
W(\Phi_i) + c\Phi_1^2 \Psi_1 + d\Phi_1^3 \Psi_2 + \tW(\tPhi_i) ,
\eeq
where the fields with negative central charge are completely decoupled from
$\tW(\tPhi_i)$. This new theory does not have a well--defined chiral ring.
However, the first three terms define a $c=0$ theory
which can be split off, leaving a nondiagonal theory with a
well--defined spectrum which is different from the spectrum of the diagonal
model.

This example suggests a simple generalization once it is put
into the proper framework.  Such a context is
 the level--rank duality which exists for all $N=2$
superconformal field theories based on hermitian symmetric coset spaces.
Consider the level 6 theory with the diagonal affine modular invariant.
Level--rank duality \cite{ks} relates the two theories
\beq
\frac{SU(2)}{U(1)}~{\Bigg |}_6 ~~
\longleftrightarrow
 ~~\frac{SU(7)}{SU(6)\times U(1)}~{\Bigg |}_1.
\lleq{minex}

Consider the potentials for these two theories. For any theory of the type
\beq
\frac{SU(m+n)}{SU(m)\times SU(n)\times U(1)}~~{\Bigg |}_1
\lleq{series}
the potential is known \cite{lvw}\cite{g} to be of the form
\beq
W= \sum_{n_1+2n_2+\cdots mn_m=m+n+1} A_{n_1\dots n_m}
                        \Phi_1^{n_1} \cdots \Phi_m^{n_m} .
\eeq
Thus the two coset models in (\ref{minex}) lead to potentials of the general
form
\bea
\frac{SU(2)}{U(1)}~{\Bigg |}_6
&:& W:= \Phi^8 \nn \\
\frac{SU(7)}{SU(6)\times U(1)}~{\Bigg |}_1
&:& W:= \Phi_1^8 + \Phi_2^4 + \Phi_2\Phi_3^2 + \Phi_4^2  \nn \\
&&  ~~~~~~~~~+~ \Phi_3 \Phi_5 + \Phi_2 \Phi_6 + \Phi_1^3 \Phi_5
                   + \Phi_1^2 \Phi_6 + \cdots \nn \\
\llea{minexpots}
where the coefficients $A_{n_1\dots n_m}$ have been dropped for convenience.

The second potential in (\ref{minexpots}) clearly combines all the potentials
discussed so far. It is important that it changes its character as one moves
around in its parameter space.
Consider e.g. the special point defined by the
superpotential
\beq
W= \Phi_1^8 + \Phi_4^2  + \Phi_2^4 + \Phi_2\Phi_3^2
                 + \Phi_3 \Phi_5 + \Phi_2 \Phi_6
 = W_1 + W_2
\eeq
where $W_1\equiv \Phi_1^8+\Phi_4^2$ and $W_2$ is the rest. Computing the
central charge of $W_2$ one finds that it vanishes! Hence it is possible
to drop this part of the potential completely. Therefore the dual potential
is, at this particular point in moduli space precisely equivalent to the
original potential.

Now move to a different point in the moduli space:
\beq
\tW =  \Phi_1^8 + \Phi_2^4 + \Phi_2\Phi_3^2 + \Phi_4^2
       + \Phi_1^3 \Phi_5 + \Phi_1^2 \Phi_6
    = \tW_1 + \tW_2
\eeq
where $\tW_1= \Phi_2^4 + \Phi_2\Phi_3^2 $ and $\tW_2$ is the rest. The
central charge of $\tW_2$ vanishes and the theory is in fact equivalent to
the potential of the level 6 minimal model with the D--invariant.
But this is precisely what is needed for mirror symmetry as I will show
in Section 4.

The construction above in fact easily generalizes to all minimal models.
Consider the level--rank duality
\beq
\frac{SU(2)}{ U(1)}~{\Bigg |}_k ~~\longleftrightarrow
\frac{SU(k+1)}{SU(k)\times U(1)}~{\Bigg |}_1.
\eeq
The potential of the minimal model is
\beq
W = \Phi_1^{k+2} + \Phi_2^2
\lleq{aminpot}
whereas the potential of the dual model is in the deformation class
\fnote{3}{This potential, as written, appears to describe the
          sum of two trivial theories with $c=0$. This is not the case
	  however, since the potential in fact does not lead to a finite
	  chiral ring at all: the theory is singular.}
\beq
\tW= \Phi_1^{k+2} + \Phi_2^{(k+2)/2} + \Phi_2\Phi_{k/2}^2 + \Phi_{(k+2)/2}^2
  + \Phi_2 \Phi_k + \Phi_{k/2} \Phi_{k/2 +2}
  + \Phi_1^2 \Phi_k + \Phi_1^{k/2} \Phi_{k/2+2} + \cdots
\eeq
The latter contains the potential (\ref{aminpot}) as well as the
Landau--Ginzburg of the minimal D--invariant
\beq
W_D= \Phi_2^{(k+2)/2} + \Phi_2\Phi_{k/2}^2
\eeq
as well as the marginal directions
\beq
\Phi_2\Phi_k ~, ~~~\Phi_{k/2}\Phi_{k/2 +2}
\eeq
and
\beq
\Phi_1^2 \Phi_k~, ~~~ \Phi_1^{k/2}\Phi_{k/2 +2}.
\eeq
It is therefore always possible to split off a trivial $c=0$ theory in the
level--rank--dual theory and obtain thereby a minimal theory in which
the diagonal affine invariant has been replaced by the D--invariant.

With the Landau--Ginzburg theories just described it is possible to analyze
a large variety of mirror pairs by tensoring other exactly solvable
$N=2$ models or, more generally, by adding some arbitrary $N=2$ supersymmetric
Landau--Ginzburg potential. For the sake of generality it should be
emphasized, however, that the idea introduced in Section 2 can be applied to
Landau--Ginzburg potentials that do not derive from the coset construction.
 Consider e.g. the pair of potentials described by
\beq
W(\Phi_i) = \Phi_1^{25} + \Phi_1\Phi_2^{16} + \Phi_3^2
\eeq
and
\beq
\tW(\tPhi_i) = \tPhi_1^{25} + \tPhi_1\tPhi_2^8 + \tPhi_2\tPhi_3^2,
\eeq
both of which are nonminimal theories, i.e. they do not belong to
ADE--type potentials. It is easy to see however that the theory described by
$\tW$ can be obtained from the theory $W$ via a $\ZZ_2$--orbifold with
respect to the action
\beq
\ZZ_2: [0~1~1]
\eeq
and a fractional transformation as described in ref. \cite{ls4}. The only new
ingredient here is the fact that the minimal part itself is coupled to another
field $\Phi_1$.

The trivialization of the theory $\tW$ leads to the $c=0$ theory
\beq
\tW'(\tPhi_i, \Psi_i) =\tW(\Phi_i) + \sum \tPhi_i\Psi_i.
\eeq
Moving along mixed directions, which exist in the case at hand, it is
possible to arrive at the potential
\beq
W(\Phi_i)+\Phi_1\Psi_1 +\Phi_1^2\Psi_2 + \Phi_1^{11}\Psi_3 +\tW(\tPhi_i).
\eeq
Splitting off the $c=0$ theory defined by the first four terms
becomes possible and leaves the new theory $\tW(\tPhi_i)$.

It is clear that the example of the last paragraph is just an iteration
of the basic construction involving the transition from the diagonal affine
invariant of the minimal model to the nondiagonal D--invariant although
dressed up a bit with some additional fields. Similarly
one can, of course, consider repeated iterations to construct ever more
complicated transititons.

There are, however, other potentials that occur in  the class of all
Landau--Ginzburg vacua \cite{alro},\cite{maha}  which are not of the
types covered by the examples above. Even though it is not, at present,
clear what precisely the
range of applicability of this new construction is, it can be shown that it is
more general than the infinite series discussed sofar. In ref. \cite{mine}
it is shown that mirror constructions involving the transposition of
the superpotential \cite{bh} can be described in this manner as well.
Furthermore it is possible to relate mirror pairs of potentials  that cannot
be related by any known means.
\vglue 0.6cm
{\bf\noindent 4. Marginal Singular Flows between Mirror Pairs of
Landau--Ginzburg \break Theories}
\vglue 0.4cm
\noindent
Consider the vacuum described by the tensor product of four minimal
$N=2$ superconformal theories $(1\cdot 6\cdot 31\cdot 86)$ endowed
with the diagonal invariant in each factor. Using the results of
\cite{m}\cite{vw} the Landau--Ginzburg potential
of this theory is described by the Fermat type polynomial
\beq
W= \Phi_1^3 + \Phi_2^{33} + \Phi_3^{88} + \Phi_4^8.
\eeq
The field theoretic limit of this model describes a particular point in the
deformation class
\beq
\IP_{(88,8,3,33,132)}[264]^{57}_{-48}
\eeq
the mirror of which can be shown, via the fractional transformations
introduced
in \cite{ls4}, to be described by the potential
\beq
W= \Phi_1^3 + \Phi_2^{33} + \Phi_3^{88} + \Phi_4^4 + \Phi_2\Phi_5^2.
\eeq
 This potential leads to the Calabi--Yau configuration
\beq
\IP_{(88,8,3,66,99)}[264]^{81}_{48}.
\eeq
The  important operation in order to produce the mirror of the ground state
thus is to simply replace the diagonal invariant in the level 6 theory by
the D--invariant. This operation leaves the first three terms in the
potential invariant. The relevant potentials to consider are
\beq
W(\Phi_i) = \Phi_4^8+\Phi_5^2
\eeq
and
\beq
\tW(\tPhi_i) = \tPhi_4^4+\tPhi_4\tPhi_5^2
\eeq
leaving the other fields as spectators.

These two polynomials, however, are just
those considered in eqs. (4,5) above and hence the previous discussion
immediately applies  to this mirror pair of Landau--Ginzburg theories:
After including the fields
$\tPsi_i$ with negative central charge the extended theory becomes
\beq
\Phi_1^3+\Phi_2^{33}+\Phi_3^{88}+
\Phi_4^8+\Phi_5^2+\tPhi_4^4+\tPhi_4\tPhi_5^2 +\tPhi_4\tPsi_4+\tPhi_5\tPsi_5.
\eeq
and moving to the singular configuration
\beq
\Phi_1^3+\Phi_2^{33}+\Phi_3^{88}+\tPhi_4^4+\tPhi_4\tPhi_5^2 +
\Phi_4^8+\Phi_5^2 + \Phi_4^2\tPsi_4 +\Phi_4^3\tPsi_5
\eeq
allows to split off the trivialization of the diagonal configuration,
leaving the mirror theory of original model.
It should be noted that instead of just reinstating the monomials defining
the original theory it is possible to also add all marginal operators
that can be constructed from the first three scaling fields. Thus it is
possible to mirror map a submanifold of the moduli space.

Furthermore it is possible to consider nondiagonal theories by adding
potentials  that do not derive from minimal $N=2$ superconformal theories.
An example of such a mirror pair is described by
\beq
\IP_{(3,11,16,10,40)}[80]^{19,27}_{-16} ~\ni ~
\{\Phi_1^{23}\Phi_2 + \Phi_1\Phi_2^7 + \Phi_3^5+\Phi_4^8+\Phi_5^2=0\}
\eeq
and
\beq
\IP_{(3,11,16,20,30)}[80]^{27,19}_{16} ~\ni ~
\{\Phi_1^{23}\Phi_2 + \Phi_1\Phi_2^7 + \Phi_3^5+\Phi_4^4+\Phi_4\Phi_5^2=0\}.
\eeq

Finally there exist, of course, also potentials which do not contain any
nontrivial minimal factor at all and the question arises whether the
concepts
 introduced in \cite{mine} also apply to such theories. This is indeed
 the case
 as the  pair of theories
\beq
\IP_{(4,6,50,15,25,50}[100]_{-8}^{33,37} ~~\ni ~~
\{\Phi_1^{25}\Phi_2 +\Phi_2^{16} +\Phi_3^2 +\Phi_4^5\Phi_5 +\Phi_5^4=0\}
\eeq
and
\beq
\IP_{(4,12,44,15,25)}[100]_8^{37,33} ~~\ni ~~
\{\tPhi_1^{25}\tPhi_2+\tPhi_1\tPhi_2^8+\tPhi_2\tPhi_3^2+
  \tPhi_4^5\tPhi_5+\tPhi_4^4=0\}
\eeq
illustrates. A more detailed discussion of these matters can be found in ref.
\cite{mine}.
\vglue 0.6cm
{\bf\noindent 5. Constructing Mirror Theories}
\vglue 0.4cm
\noindent
So far I have always assumed that a pair of theories is given a priori
and then attempted to
construct a flow from one to the other. The question arises whether it is
possible to let the construction determine to which theory the deformed
model wants to flow. This is a somewhat involved problem which I will discuss
only very briefly in the context of the simplest theory encountered above.

Consider again the potential (\ref{min6d}) with central charge $c=9/4$ as
starting point. The fields $\Psi_i$ that appear in the trivialized theory
have charges $3/4,5/8$ respectively. We are interested in all potentials
$W(\Phi_1,\Phi_2)$ which satisfy a number of constraints:
\begin{itemize}
\item  $W(\Phi_1,\Phi_2)$ describes a theory of central charge $c=9/4$
\item  Marginal operators of the form $\Phi_i^{a_i}\Psi_j$ have to exist
\item  $W(\Phi_1,\Phi_2)$ must have an isolated singularity at the origin.
\end{itemize}
In the case at hand there are only a few possible operators that
can appear. (I) The operators can be of the form
$\Phi_1^{a_1} \Psi_1, \Phi_2^{a_2} \Psi_2$. The central charge condition
then dictates that the only possible solution is $a_1=1=a_2$ which are
the operators used to trivialize the original theory.
(II) The operators are of the form $\Phi_1^{a_1} \Psi_1,~\Phi_1^{a_2} \Psi_2$,
in which case $2a_2=3a_1$. Thus $a_1$ must be even $a_1=2n$ and $q_1=1/8n$ and
 $q_2=\frac{1}{8}(5-\frac{1}{n})$. The requirement that the potential
 $W(\Phi_1,\Phi_2)$ has
 an isolated singularity only at the origin finally determines $n=1$ and
 hence we have {\it derived} from the trivialized theory (\ref{min6d}) the
 potential (\ref{min6a}) dressed up with a trivial factor.
\vglue 0.6cm
{\bf \noindent 5. Acknowledgements \hfil}
\vglue 0.4cm
\noindent
I'm grateful to  Per Berglund, Wolfgang Lerche and Jan Louis for discussions.
\vglue 0.6cm
{\bf\noindent 6. References \hfil}
\vglue 0.4cm

\end{document}